\documentclass[%
 reprint,
 superscriptaddress,
 amsmath,amssymb,
 prl,aps
]{revtex4-2}

\usepackage{graphicx}
\usepackage{dcolumn}
\usepackage{bm}
\usepackage{braket}
\usepackage{color}
\usepackage{hyperref}
\usepackage{orcidlink}

\newcommand{\schro}{Schr\"odinger}

\def\apj{ApJ}
\def\mnras{MNRAS}

\def\araa{ARA\&A}

\def\jcap{Journal of Cosmology and Astroparticle Physics}

\begin{document}

\title{Diverse dark matter haloes in Two-field Fuzzy Dark Matter}

\author{Hoang Nhan Luu\,\orcidlink{0000-0001-9483-1099}}
\email{hoang.luu@dipc.org}
\affiliation{Donostia International Physics Center, Basque Country UPV/EHU, San Sebastian, E-48080, Spain}
\affiliation{Department of Physics, Kavli Institute for Astrophysics and Space Research, Massachusetts Institute of Technology, Cambridge, MA 02139, USA}

\author{Philip Mocz\,\orcidlink{0000-0001-6631-2566}}
\email{pmocz@flatironinstitute.org}
\affiliation{Flatiron Institute, 162 5th Ave, New York, NY, 10010, USA}

\author{Mark Vogelsberger\,\orcidlink{0000-0001-8593-7692}}
\email{mvogelsb@mit.edu}
\affiliation{Department of Physics, Kavli Institute for Astrophysics and Space Research, Massachusetts Institute of Technology, Cambridge, MA 02139, USA}

\author{Alvaro Pozo\,\orcidlink{0009-0006-1992-0722}}
\affiliation{Donostia International Physics Center, Basque Country UPV/EHU, San Sebastian, E-48080, Spain}
\affiliation{University of the Basque Country UPV/EHU, Department of Theoretical Physics, Bilbao, E-48080, Spain}

\author{Tom Broadhurst\,\orcidlink{0000-0002-8785-8979}}
\affiliation{Donostia International Physics Center, Basque Country UPV/EHU, San Sebastian, E-48080, Spain}
\affiliation{University of the Basque Country UPV/EHU, Department of Theoretical Physics, Bilbao, E-48080, Spain}
\affiliation{Ikerbasque, Basque Foundation for Science, Bilbao, E-48011, Spain}

\author{S.-H. Henry Tye\,\orcidlink{0000-0002-4386-0102}}
\affiliation{Department of Physics and Institute for Advanced Study, The Hong Kong University of Science and Technology, Hong Kong}
\affiliation{Department of Physics, Cornell University, Ithaca, NY 14853, USA}

\author{Tao Liu\,\orcidlink{0000-0002-5248-5076}}
\affiliation{Department of Physics and Institute for Advanced Study, The Hong Kong University of Science and Technology, Hong Kong}

\author{Leo W.H. Fung\,\orcidlink{0000-0002-5899-3936}}
\affiliation{Department of Physics and Institute for Advanced Study, The Hong Kong University of Science and Technology, Hong Kong}

\author{George F. Smoot\,\orcidlink{0000-0001-7575-0816}}
\affiliation{Donostia International Physics Center, Basque Country UPV/EHU, San Sebastian, E-48080, Spain}
\affiliation{Department of Physics and Institute for Advanced Study, The Hong Kong University of Science and Technology, Hong Kong}
\affiliation{Paris Centre for Cosmological Physics, APC, AstroParticule et Cosmologie, Universit\'e de Paris, CNRS/IN2P3, CEA/lrfu, 10, rue Alice Domon et Leonie Duquet, 75205 Paris CEDEX 13, France emeritus}
\affiliation{Physics Department, University of California at Berkeley, CA 94720, Emeritus}

\author{Razieh Emami\,\orcidlink{0000-0002-2791-5011}}
\affiliation{Center for Astrophysics, Harvard \& Smithsonian, 60 Garden Street, Cambridge, MA 02138, USA}

\author{Lars Hernquist\,\orcidlink{0000-0001-6950-1629}}
\affiliation{Center for Astrophysics, Harvard \& Smithsonian, 60 Garden Street, Cambridge, MA 02138, USA}

\date{\today}

\begin{abstract}
Fuzzy dark matter (FDM) is a compelling candidate for dark matter, offering a natural explanation for the structure of diffuse low-mass haloes. However, the canonical FDM model with a mass of $10^{-22}~{\rm eV}$ encounters challenges in reproducing the observed diversity of dwarf galaxies, except for possibly scenarios where strong galactic feedback is invoked. The introduction of multiple-field FDM can provide a potential resolution to this diversity issue. The theoretical plausibility of this dark matter model is also enhanced by the fact that multiple axion species with logarithmically-distributed mass spectrum exist as a generic prediction of string theory. In this paper, we consider the axiverse hypothesis and investigate non-linear structure formation in the two-field fuzzy dark matter (2FDM) model. Our cosmological simulation with an unprecedented resolution and self-consistent initial conditions reveals the diverse structures of dark matter haloes in the 2FDM model for the first time. Depending on the formation time and local tidal activities, late-time haloes can host solitons of nested cores or solitons of one dominant species.
\end{abstract}

\maketitle

\paragraph{Introduction.} 

Cosmology in the presence of multiple axions, namely ``the axiverse", is generically expected in string theory~\cite{Svrcek:2006yi, Arvanitaki:2009fg}. Predictions for the axion mass spectrum depend on the details of the precise dynamics of the dimensional compactification that describes our universe, which so far remains unknown. Knowing that axion masses are generated only by non-perturbative (instanton) effects, typical axion masses are exponentially smaller than the standard model scale, and the axion spectrum covers a wide range in masses, which may even reach as low as the Hubble scale of $10^{-33}~{\rm eV}$. The fuzzy dark matter (FDM) proposal composed of ultralight axions of mass $\sim 10^{-22}$ eV~\cite{Hu:2000ke, Marsh:2015xka, Hui:2016ltb, Ferreira:2020fam} is well motivated from this perspective.
 
Cosmological simulations of FDM reveal rich wave-like structures of constructive and destructive interference on the de Broglie scale of $\sim 1~{\rm kpc}$, including the NFW-like haloes of galaxies and a stable ``soliton'' ground state at the core of every collapsed structure~\cite{Schive:2014dra, Schwabe:2016rze, May:2021wwp, Mocz:2017wlg}. This soliton formation and wave-like behaviour are unique properties that distinguish FDM from cold dark matter (CDM)~\cite{Mocz:2019pyf, Mocz:2019uyd} and provide an alternative solution to the small-scale issues of the standard $\Lambda$CDM model~\cite{Klypin:1999uc, BoylanKolchin:2011, deBlok:2010, Bullock:2017xww}, without invoking the sub-grid physics of ``baryonic feedback''~\cite{Governato:2012, DiCintio:2013qxa, Vogelsberger:2019ynw} that is yet to be understood.

However, the canonical model of FDM faces several challenges from observational Lyman-$\alpha$ data, which excludes either the particle mass up to $10^{-20}~{\rm eV}$~\cite{Armengaud:2017nkf, Irsic:2017yje, Rogers:2020ltq} or the cosmological abundance to less than $30\%$~\cite{Kobayashi:2017jcf}. Stellar dynamics inside galaxies and satellites also places stringent constraints on the mass spectrum~\cite{Chen:2016unw, Marsh:2018zyw, Broadhurst:2019fsl, Safarzadeh:2019sre, Hayashi:2021xxu, Dalal:2022rmp}. What has become increasingly clear is that FDM haloes are not able to simultaneously fit two distinguishable classes of dwarf galaxies, namely the classical dwarf spheroidals (dSphs)~\cite{Walker:2009} and the physically much smaller and less luminous of ultra-faint dwarfs (UFDs)~\cite{Simon:2019}. The primary concern is that star clusters in UFDs tend to be overheated by density fluctuations from the soliton or interference granules~\cite{Marsh:2018zyw, Dalal:2022rmp}, unless they are significantly stripped away via tidal evolution~\cite{Schive:2019rrw, Chiang:2021uvt}. This paradoxical situation of FDM is reminiscent of the diversity problem in galactic rotation curves encountered in the standard CDM paradigm~\cite{Oman:2015xda}.

Meanwhile, it has been recently suggested that a cosmological model with multiple ultralight axions, or equivalently multiple FDM species~\cite{Luu:2018afg, Huang:2022ffc, Gosenca:2023yjc}, may accommodate diverse profiles of dark matter haloes~\cite{Guo:2020tla, Eby:2020eas, Maleknejad:2022gyf, Luu:2023dmi, vanDissel:2023vhu}, hence circumvents the aforementioned issue with dwarf galaxies. The two-field fuzzy dark matter (2FDM) model with particle masses separated by 1-2 orders of magnitude provides the most simplified construction of the axiverse where ``halo diversity'' is anticipated~\cite{Luu:2018afg, Pozo:2023zmx}. Specifically, density fluctuations in separate patches of the universe with varying fractions of the 2FDM species would form haloes of different inner core profiles at late times. Despite the feasibility of this idea, cosmological simulations for the 2FDM model have not been able to capture the relatively wide range of de Broglie wavelengths spanned by both FDM species due to the enhanced dynamical range required by a large particle mass hierarchy.
 
In this paper, we make the first attempt to simulate the structure formation for two axion species differing in particle mass by a factor of 5, the highest ratio by far for cosmological simulations of this type. We develop an optimized spectral solver for 2FDM equations of motion with self-consistent initial conditions solved from 2FDM linear perturbations. The improved simulation technique combined with the large mass factor allows$-$for the first time$-$investigation of dark matter haloes and their diverse structures with unprecedented resolution.

\paragraph{Cosmological simulation.}

In the non-relativistic limit, the dynamics of the 2FDM model is governed by the coupled \schro-Poisson (SP) equations, which can be written explicitly in the comoving coordinates as
\begin{equation}
\begin{aligned}
    &i\hbar\dfrac{\partial\psi_1}{\partial t} = -\dfrac{\hbar^2}{2m_1a^2} \nabla^2\psi_1 + \dfrac{m_1}{a} \Phi\psi_1, \\
    &i\hbar\dfrac{\partial\psi_2}{\partial t} = -\dfrac{\hbar^2}{2m_2a^2} \nabla^2\psi_2 + \dfrac{m_2}{a} \Phi\psi_2, \\
    &\nabla^2\Phi = 4\pi G \left(|\psi_1|^2 + |\psi_2|^2 - \bar{\rho} \right).
\end{aligned} \label{Eq:schro-equations}
\end{equation}
Here $\psi_1$ and $\psi_2$ represent wavefunctions of the 2FDM fields with a mass $m_1$ and $m_2$, respectively. $\Phi$ denotes a gravitational potential sourced by the total density, $\rho = \rho_1 + \rho_2 \equiv |\psi_1|^2 + |\psi_2|^2$. $\bar{\rho}$ is the average density over the comoving volume of interest and $a$ is the scale factor. For clarity reasons, let us refer to the light field and the heavy field as $\psi_1$ and $\psi_2$ from now on, assuming $m_2 > m_1$. We will also refer to the total field as the sum of $\psi_1$ and $\psi_2$ in terms of dark matter density.

The SP equations in \eqref{Eq:schro-equations} can be solved numerically by the pseudo-spectral method which evolves the system unitarily via a series of ``kick'' and ``drift'' steps as described in Ref.~\cite{Luu:2023dmi}. To achieve desired performance and scalability, we develop a fully parallelized solver optimized for the 2FDM model in this work.

We perform a high-resolution cosmological simulation in the 2FDM model where $m_1 = 10^{-22}~{\rm eV}$ and $m_2 = 5 \times 10^{-22}~{\rm eV}$, with a density ratio of $\beta_2 \equiv \Omega_2/\Omega_m = 0.7$. Although the chosen values of $m_1$ and $m_2$ are in tension with observations, the relevant physics still applies for higher particle masses of the same ratio $m_2/m_1$ thanks to the scale invariance of SP equations. The simulation volume has a side length of $1.7~{\rm Mpc}/h$ in each dimension. The spectral resolution is $2048^3$ for the 2FDM fields. Periodic boundary condition is automatically applied in the spectral solver. The background cosmology is set up with cosmological parameters from the most updated Planck data~\cite{Planck:2018vyg} except for $A_s = 10^{-8}$. This enhanced value of $A_s$ is to compensate for the density fluctuations from large-scale modes in a small simulation volume.

To obtain self-consistent initial conditions, at first we employ \texttt{N-GenIC}~\cite{Springel:2005} to generate a realization of initial random phases in the momentum space. The initial power spectra are then computed with the Boltzmann solver $\texttt{CAMB}$~\cite{Lewis:1999bs} specifically modified for the 2FDM model. Finally, the initial wavefunctions of the 2FDM fields are solved from Madelung (fluid) equations. More details about how the initial conditions are set up can be found in the Supplemental Material.

We evolve the 2FDM system from the starting redshift of $z = 127$ to the final redshift of $z = 3.4$. In terms of convergence, we find that some smallest features are slightly under-resolved at the final redshift, but they do not affect the main results discussed below.

\begin{figure}
    \centering
    \includegraphics[scale=0.5]{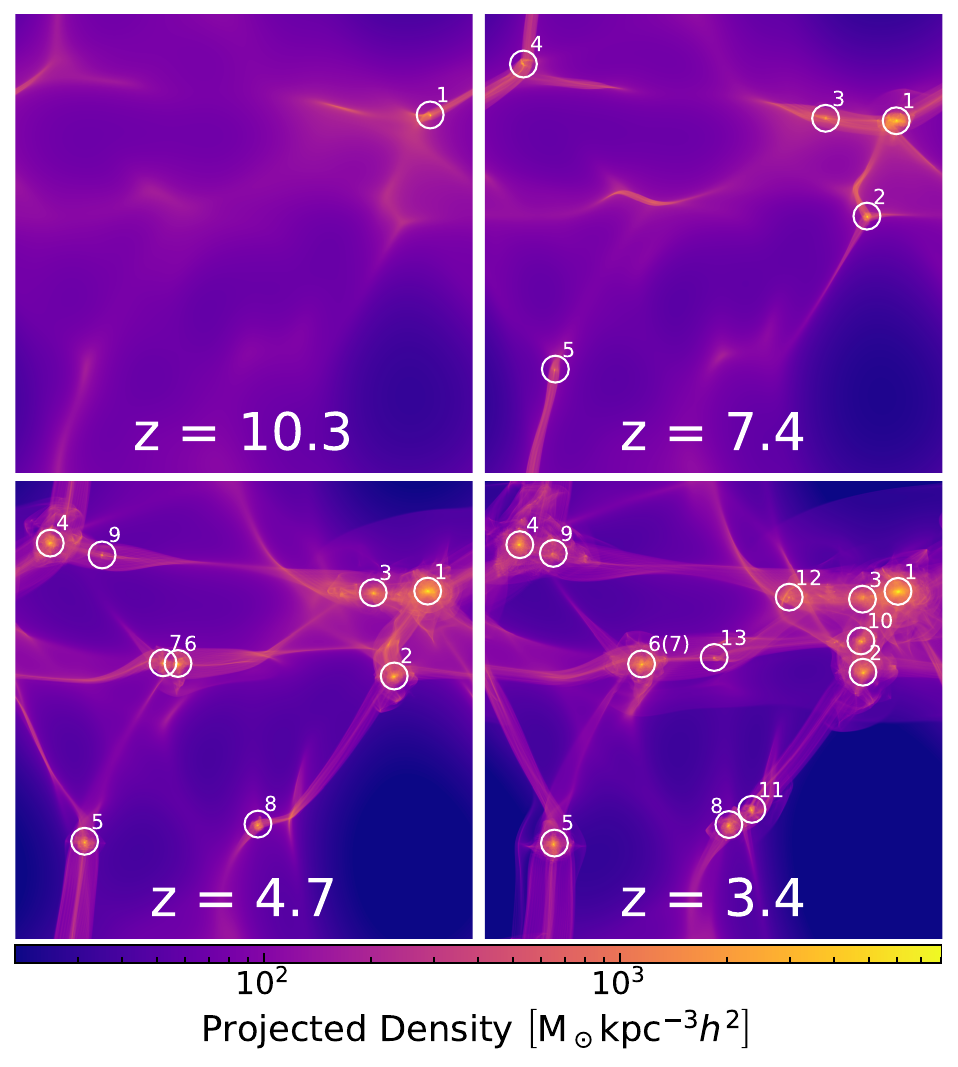} \\
    \caption{Projected densities of the total field in a simulation volume with a side length of $1.7~{\rm Mpc}/h$ at several redshifts. The white circles mark all haloes formed with a  solitonic structure at the corresponding redshift denoted in each panel. The number adjacent to each halo indicates (approximately) its formation order. We note that Halo 6 and 7 have merged at the last redshift.}
    \label{Fig:proj_evolve}
\end{figure}

\begin{figure}
    \centering
    \includegraphics[scale=0.78]{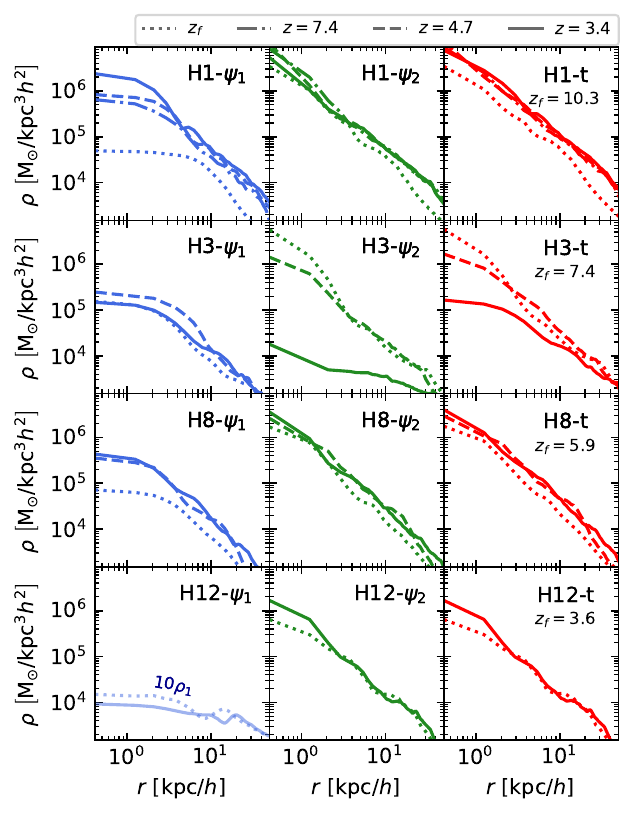}
    \caption{Radial profiles of some representative haloes at several redshifts. In each panel, $z_f$ denotes the first redshift at which the corresponding halo forms with a solitonic structure in the simulation. Blue, green and red curves correspond to profiles of $\psi_1, \psi_2$ and the total field. H$x-\psi_1$, H$x-\psi_2$ and H$x-$t denote $\psi_1, \psi_2$ and the total-field profiles of Halo $x$, respectively. In each panel, the halo center is chosen as the barycenter of the total field. Since the density of $\psi_1$ in Halo 12 is too low, its profiles (in faded blue) are multiplied by 10 for better illustration.}
    \label{Fig:rad_evolve}
\end{figure}

\paragraph{Structure formation in the 2FDM cosmology.} 

Fig.~\ref{Fig:proj_evolve} shows how structures evolve in the 2FDM cosmology. We observe in total 13 haloes formed by the end of our simulation. These haloes are numbered based on their (approximate) formation history where smaller numbers indicate haloes that form earlier.

To understand the halo evolution, we compare the radial profiles of a few haloes from their formation redshift ($z = z_f$) to the final redshift ($z = 3.4$) in Fig.~\ref{Fig:rad_evolve}. Only four haloes are displayed here as they represent different viable evolution patterns found among the other ones. Generally, every halo starts off with a density fraction of $\psi_2$ higher than that of $\psi_1$ as expected from their initial cosmological abundance. Each halo, however, evolves to a distinct final state depending on its formation time.

In a few haloes that form first in the simulation, such as Halo 1, $\psi_2$ reaches a stable configuration shortly while the density of $\psi_1$ keeps growing until $z = 3.4$. As a result, $\psi_1$ becomes comparable to $\psi_2$ in terms of mass and density content at the last redshift. Interestingly, Halo 1 is just a typical example to show that the higher cosmological density of $\psi_2$, {\it i.e.}, $\beta_2 > 0.5$, does not always translate to its dominance inside individual haloes.

In other haloes that form at later redshifts, such as Halo 8, the 2FDM fields experience a similar growth but $\psi_1$ ends up a sub-dominant component compared to $\psi_2$. It seems that the distribution of $\psi_2$ in Halo 8 is not massive enough to support the clustering of $\psi_1$. In addition, the mass accumulation rate of $\psi_1$ here is relatively slow, which means $\psi_1$ might have already settled in its virialized state without further growth.

There are also extreme cases such as Halo 12 which is severely deficient of $\psi_1$. The lack of $\psi_1$ in this halo can be explained by its late formation time ($z_f = 3.6$). Since $\psi_1$ only accounts for $30\%$ of the total DM budget, most of $\psi_1$ abundance has already clustered in haloes that form earlier. Thus, we anticipate that any haloes forming later than Halo 12 are also completely dominated by $\psi_2$.

Halo 3 is somewhat special as it undergoes a separate track from the remaining haloes. Since this halo locates in the neighborhood of another massive object, {\it i.e.}, Halo 1, it evolves under a gravitational potential during the entire lifetime. Thus, we observe strong tidal disruption in the density profile of $\psi_1$ and $\psi_2$ when Halo 3 approaches Halo 1. Most notably, only the mass of $\psi_2$ is significantly stripped away while most of $\psi_1$ mass remains intact. The reason why there exists such an asymmetry is unknown at the moment, but this phenomenon indicates a possible mechanism to create $\psi_1$-dominated haloes.

Another curious scenario is the coalescence of Halo 6 and 7 into a more massive halo, as seen in Fig.~\ref{Fig:proj_evolve}. Halo mergers such as this pair are expected to be common in a larger volume. Here the evolution of $\psi_1$ and $\psi_2$ similarly follow those of Halo 1, {\it i.e.}, they also yield an equivalent amount of $\psi_1$ and $\psi_2$ at the end.

\paragraph{Diverse structures of 2FDM haloes.}

As the virialized configuration of haloes are not universal in a 2FDM cosmology, it is important to examine the demographics of each halo population in more details.

\begin{figure*}
    \centering
    \includegraphics[scale=0.6]{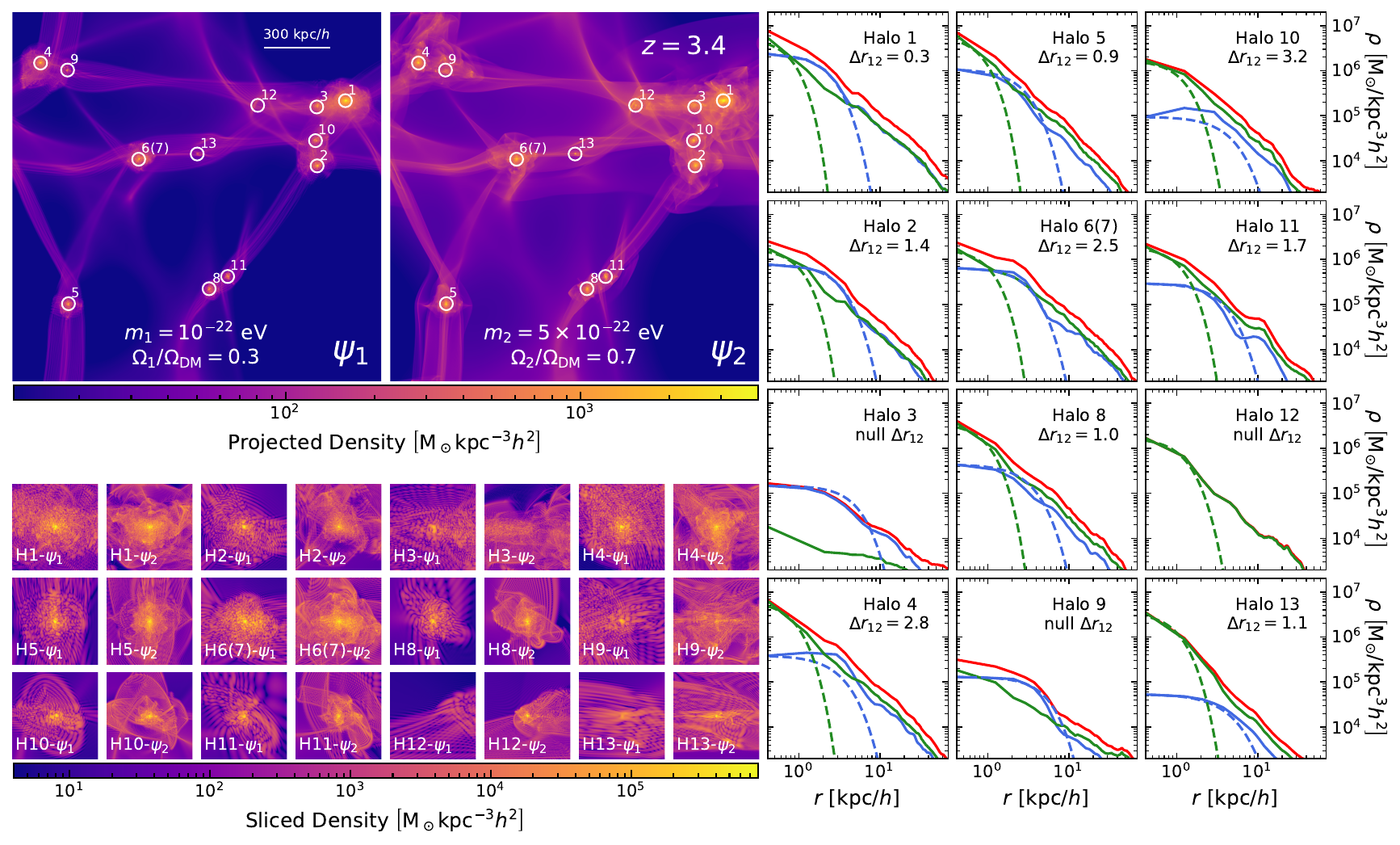}
    \caption{(Left top) Projected densities of $\psi_1$ and $\psi_2$ at $z = 3.4$. The white circles mark the haloes found at this redshift, as in Fig.~\ref{Fig:proj_evolve}. (Left bottom) Sliced densities of the haloes marked in the above panels. Each slice displays a cross section through the center of the associated halo with a side length of $200~{\rm kpc}/h$. (Right) Radial profiles of all haloes at $z = 3.4$. The solid curves show the simulation profiles of the total field (red), $\psi_1$ (blue), and $\psi_2$ (green). The dashed curves show the best-fit soliton profiles of $\psi_1$ (blue) and $\psi_2$ (green) in each halo. As the barycenters of $\psi_1$ and $\psi_2$ are not always aligned, $\Delta r_{12}$ denotes the distance between them in the units of ${\rm kpc}/h$. The soliton profile of only one field is shown in Halo 3, 9 and 12 because the other field does not host a soliton. $\Delta_{12}$ is also undefined for these haloes.} 
    \label{Fig:proj}
\end{figure*}

Fig.~\ref{Fig:proj} provides a complete landscape of the comoving volume and 12 haloes found at $z = 3.4$. The projected densities (top-left panels) show an almost identical filamentary structure of $\psi_1$ and $\psi_2$ on large scales and the lower density distribution of $\psi_1$ compared to the one of $\psi_2$. The sliced densities (bottom-left panels) give a close-up view of individual haloes with their central solitons surrounded by density granules from wave interference. Most important of all, the radial profiles (right panels) clearly illustrate a diversity of haloes with distinct solitonic core structures. We find that 2FDM haloes can be separated into three populations as follows.

The first population consists of Halo 1, 2, 4, 5 and 6(7). The radial profiles of these haloes show a high fraction of $\psi_1$ compared to $\psi_2$ with the presence of two central cores. In Halo 1, 2, 5 where the cores of $\psi_1$ and $\psi_2$ are approximately concentric ($\Delta r_{12} < 1.5~{\rm kpc}$), the simulation profiles perfectly match the so-called \textit{nested soliton}, namely the ground-state solution of the time-independent SP equations~\cite{Luu:2018afg, Luu:2023dmi}. On the contrary, in Halo 4 and 6(7) where the two cores are not aligned ($\Delta r_{12} > 2~{\rm kpc}$) due to soliton random walk~\cite{Schive:2019rrw}, the nested soliton does not yield a good fit as expected. Overall, the haloes belonging to this population would be observed as ones with centrally nested structures (from the total field perspective).

The second population consists of Halo 8, 10, 11, 12 and 13. These haloes include an extremely low to moderate amount of $\psi_1$. In this case, even when the two cores of $\psi_1$ and $\psi_2$ form in a concentrically nested configuration, only the soliton of $\psi_2$ is visible as it overwhelmingly dominates the one of $\psi_1$. As the outer regions of these haloes are also dominated by $\psi_2$, they would be most likely observed as $\psi_2$-only haloes.

The third population consists of Halo 3 and 9. As previously mentioned, Halo 3 has been subject to tidal interactions since its formation. This effect causes a complete disruption of the $\psi_2$ soliton, so that Halo 3 is eventually dominated by $\psi_1$ with its soliton. A similar phenemenon also occurs with Halo 9 during its evolution under the gravitational potential of Halo 4 (see Fig.~\ref{Fig:proj_evolve}), resulting in a considerable mass loss of $\psi_2$. Although the density of $\psi_2$ is still comparable to the one of $\psi_1$ here, there is no $\psi_2$ soliton formed in this halo (see the sliced and projected densities). Eventually, Halo 3 and 9 would be observed as $\psi_1$-only haloes.

\paragraph{Halo diversity in observation.}

Can we seek the aforementioned halo populations in observational data? Previous studies~\cite{Luu:2018afg, Pozo:2023zmx} suggested that the DM profiles of dSphs and UFDs can be explained by two axion species with $m_1 \sim 10^{-22}~{\rm eV}$ and $m_2 \sim 10^{-20}~{\rm eV}$, respectively.

If we assume that the current simulation results can be extrapolated to a larger value of $m_2$, the $\psi_1$-only and $\psi_2$-only populations mentioned above would satisfy the observational constraints of dSphs and UFDs, respectively. However, there are two problems with this identification. Firstly, the $\psi_2$-only haloes are more common than the $\psi_1$-only ones, but the number of presently detected UFDs are of the same order of magnitude with dSphs. Secondly, $\psi_2$-only haloes generally form later than $\psi_1$-only haloes whereas UFDs are believed to be much older than dSphs~\cite{Simon:2019}. It is possible that UFDs should be identified with extremely-low-$\psi_1$ haloes like Halo 12, which would appear much earlier in a higher-$m_2$ 2FDM simulation. However, we need cosmological simulations with a proper implementation of star formation and baryonic physics to address these problems. At the moment there exist feasible but yet definitive connections between our simulation haloes and the observed dwarf galaxies.

On the other hand, the first population with nested soltions can be identified with normal galaxies such as Milky Way (MW). There are, in fact, some observational evidence of such nested profile in the MW center. For instance, Ref.~\cite{DeMartino:2018zkx} found that the excess velocity dispersion of the central MW bulge stars could be accounted for by a $\psi_1$ soliton of a mass $10^9~{\rm M}_\odot$. Meanwhile, a similar analysis of FDM in the nuclear star cluster of MW~\cite{Toguz:2022} found a positive hint for a soliton of an FDM field with a mass $10^{-20.5}~{\rm eV}$, which is approximately the boson mass expected for $\psi_2$. As such, we may hope to continue searching for nested solitons in other galaxies in the near future.

The existence of $\psi_1$-deficient galaxies such as Halo 12 for a 2FDM cosmology with non-negligible $\Omega_1$ also affects the existing limits on canonical FDM derived from data of isolated systems~\cite{Dalal:2022rmp, Powell:2023jns}. Since the considered system may be hosted by a 2FDM halo like this one, their constraints should apply to the particle mass of the heavier species of FDM, ${\it i.e.}$, $m_2$, instead of $m_1$ as the most conservative estimation. In addition, it is necessary to revisit these constraints with tidal stripping taken into account because the stellar heating effect caused by transient density fluctuations generic to FDM are highly suppressed in that context~\cite{Schive:2019rrw, Chiang:2021uvt}.

\paragraph{Conclusion and outlook.}

We have demonstrated that the 2FDM cosmology can provide rich and complex structure formation, with the diversity of dark matter haloes being one of its most distinguishing features. Most young and small haloes are dominated by $\psi_2$ and they only see $\psi_2$ solitons as the central major components. Old and more massive haloes, on the other hand, incorporate comparable amounts of $\psi_1$ and $\psi_2$. Hence, these haloes host solitons with observable nested structures. Lastly, haloes dominated by $\psi_1$ may emerge via tidal disruption in natural encounters with nearby haloes.

It is important to note that these results can be subject to some limitations of our simulation. For instance, the major restriction of the spectral method (uniform resolution) is that we can only evolve the system to a certain point before features become smaller than what can be represented by the maximum wavenumber in the simulation. As a consequence, the peaked (central) density of $\psi_2$ in each halo can be underestimated when its soliton is not completely resolved. This drawback, however, does not affect any qualitative conclusions about halo diversity. Another caveat is that the above results only apply to the 2FDM system where $m_2/m_1 = 5$ and $\beta_2 = 0.7$. Even though we also examine simulations with other input parameters of $m_2/m_1$ and $\beta_2$, dark matter haloes here are typically dominated by a single population of haloes, {\it i.e.}, no halo diversity is realized. Extended discussions on these scenarios can be found in the Supplemental Material.

With the first compelling evidence for halo diversity clearly shown in this study, future work could aim at simulating larger cosmological volumes with higher resolution at lower redshifts, if possible. The string axiverse, as represented by the 2FDM model, still holds surprisingly many yet-to-be-discovered potentials, which will hopefully open a new window to the long-standing dark matter mystery of our time.

\paragraph{Acknowledgments.}

HN appreciates Dr.~Jens Stücker, Prof.~Raúl Angulo and Prof.~Simon White for valuable discussions and insights to the final results of this paper. HN also thanks Prof.~Raúl Angulo and Donostia International Physics Center (DIPC) for providing a simulation storage on DIPC Supercomputing Center. HN, TB, HT, TL and GS acknowledge support from the Research Grants Council of Hong Kong through the Collaborative Research Fund C6017-20G. LH acknowledges support by the Simons Collaboration on ``Learning the Universe". The simulations in this work were run on the Engaging cluster sponsored by Massachusetts Institute of Technology (MIT). TB is supported by the Spanish grant PID2023-149016NBI00 (funded by MCIN/AEI/10.13039/501100011033 and by "ERDF A way of making Europe".

\appendix

\section{Supplemental Material}

The supplemental materials include information that may be useful for readers interested in more aspects of the 2FDM model. In Sec.~A we provide the detailed derivation of the initial conditions in the 2FDM cosmology. In Sec.~B we compare and discuss 2FDM simulations with a variety of input parameters (other than the one in the main text).

\renewcommand{\thefigure}{A}
\begin{figure*}
    \centering
    \includegraphics[scale=0.65]{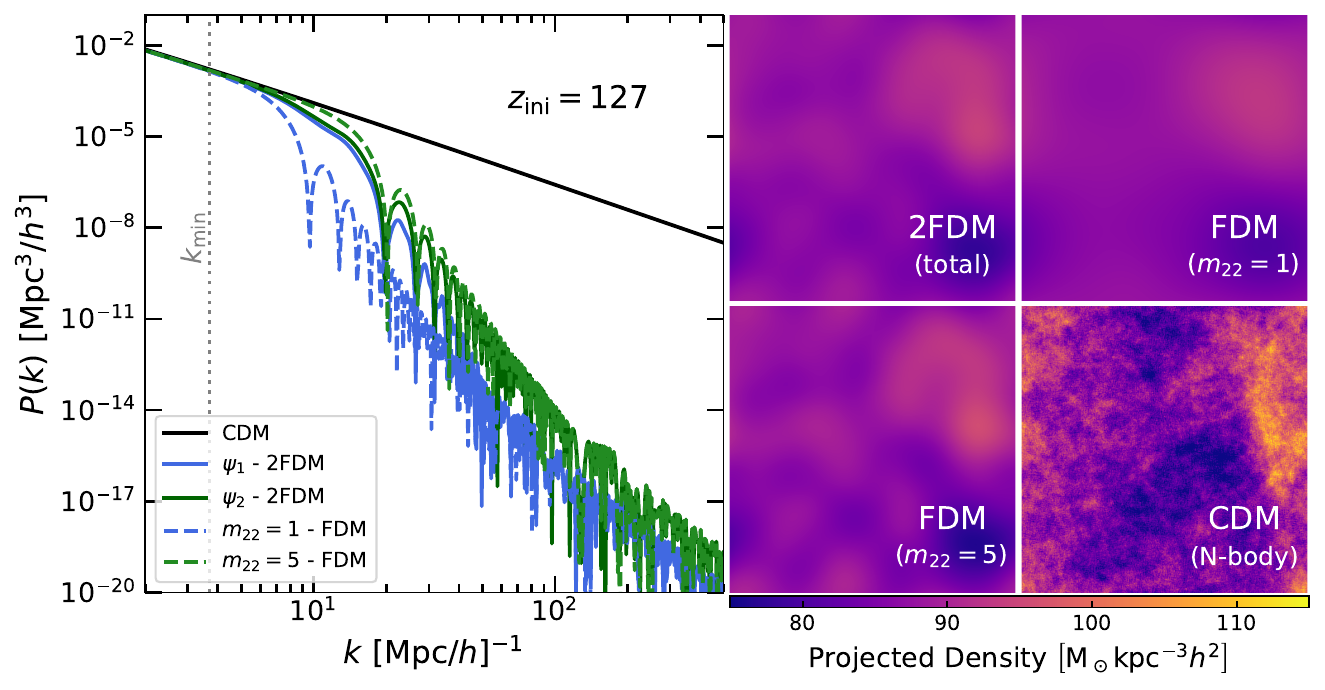} \\
    \caption{(Left) Linear power spectra at the initial redshift $z_{\rm ini} = 127$. The spectra of $\psi_1$ and $\psi_2$ in the 2FDM model ($\beta_2 = 0.7, m_1 = 10^{-22}~{\rm eV}, m_2 = 5 m_1$) are shown as the blue-solid and green-solid curves. The spectra of a light ($m_{22} = 1$) and heavy ($m_{22} = 5$) species in the FDM model are shown as the blue-dashed and green-dashed curves. The spectrum of CDM is also shown as the black-solid curve for comparison. The dotted vertical line indicates the minimum wavenumber (the largest scale) accommodated by the simulation box. (Right) Projected densities of initial fluctuations with matching phases in various dark matter models. The 2FDM/FDM initial conditions are generated as wavefunctions on a uniform grid, while the CDM ones are generated as N-body particles with \texttt{N-GenIC}~\cite{Springel:2005}.}
    \label{Fig:ini_spec}
\end{figure*}

\section{A. Initial conditions} \label{App:ics}

In single-field FDM simulations, the matter power spectrum is typically derived using a numerical Boltzmann solver such as \texttt{axionCAMB}~\cite{Hlozek:2014lca, Grin:2022} or using the approximate axion transfer function (at $z = 0$) given by~\cite{Hu:2000ke}
\begin{align}
    T (k) = \sqrt{\dfrac{P_{\rm FDM}}{P_{\rm CDM}}} = \dfrac{\cos x^3_J}{1 + x^8_J}, \label{Eq:transfer}
\end{align}
where $x_J = 1.61 m_{22}^{1/18} k/k_{J, {\rm eq}}; k_{J, {\rm eq}} = 9 m_{22}^{1/2}\,{\rm Mpc}^{-1}$ and $m_{22} = m/10^{-22}~{\rm eV}$. We note that the transfer function \eqref{Eq:transfer} assumes the total dark matter budget composed of axions and has a scale-dependent growth.

By far, FDM initial conditions are usually derived from the N-body particle distribution via density assignment such as the cloud-in-cell algorithm because it is particularly convenient to generate N-body initial conditions via publicly available codes~\cite{Springel:2005, Hahn:2013}, even for an arbitrary power spectrum. However, this approach is rather unreliable for ICs generation of field-based simulations for two reasons. Firstly, the velocities of N-body particles are typically computed with the Zel'dovich approximation or the second-order Lagrangian Perturbation Theory that is tailored for CDM perturbations. Secondly, particle discreteness might introduce abnormal mesh noise on the smallest scales of FDM simulations when being converted to fields. These effects might induce artificial formation of compact objects at late times, especially for FDM with a high particle mass. As such, the most robust and self-consistent approach is to derive the initial conditions of FDM directly from its linear perturbations in the early universe, which is implemented as follows.

For a set of random phases $\{ \theta_{\bf k} \}$ in the momentum space, the initial density fluctuations of the FDM field can be solved via an inverse Fourier transform
\begin{align}
    \delta ({\bf x}) \leftarrow {\rm ifft}\left[ \Delta_k e^{i \theta_{\bf k}} \right], \quad \Delta_k^2 = k^3 P_k/(2\pi)^3.
\end{align}
Here $P_k$ is computed at the initial redshift, {\it e.g.}, $z = 127$ in our simulation, which is then related to the present spectrum of CDM by the linear growth factor and the transfer function, $P_k = D^2(z_{\rm ini})/D^2(0) T^2(k) P_{\rm CDM}$. This procedure is then repeated to find the (conformal) time derivatives of the density field $\delta'({\bf x}) = \partial\delta/\partial\eta$, which are analogous to the particle velocities in N-body simulations. Both $\delta ({\bf x})$ and $\delta'({\bf x})$ are necessary degrees of freedom to derive the complex-valued wavefunction of the FDM field in the \schro~equation
\begin{align}
    i\hbar\dfrac{\partial\psi}{\partial t} = -\dfrac{\hbar^2}{2ma^2} \nabla^2\psi + \dfrac{m}{a} \Phi\psi \, . \label{Eq:schro-equation}
\end{align}
If we decompose $\psi = \psi_r e^{i\alpha}$ and define the Madelung velocity as $v_M \equiv \hbar/m \nabla \alpha$, the continuity equation yields a relation of $\alpha$ and $\delta'$ at the first perturbative order as
\begin{align}
    \dfrac{\partial\rho}{\partial t} + \nabla \cdot (\rho v_M) = 0 \quad\rightarrow \quad \nabla^2 \alpha = - \dfrac{m}{\hbar} a \delta' \, . \label{Eq:wave-phase}
\end{align}
We then solve Eq.~\eqref{Eq:wave-phase} to obtain the phase $\alpha$ while the modulus $\psi_r$ can be easily inferred from the density contrast by $\rho = \Omega_{\rm DM} \rho_{\rm cr} (1 + \delta) = |\psi_r|^2$, which makes the wavefunction $\psi$ fully specified.

These computations can be generalized for two axion species in the 2FDM model, providing $\delta_1, \delta_2$ and $\delta'_1, \delta'_2$. However, the transfer function of the two axion species is no longer given by \eqref{Eq:transfer}. Instead, we need to solve the linear perturbation equations of both fields
\begin{equation}
\begin{aligned}
    \delta'_i &= - ku_i - h'/2 - 3\mathcal{H}c_{s,i}^2\delta_i - 9\mathcal{H}^2c^2_{s,i}u_i/k \, , \\
    u_i' &= -\mathcal{H}u_i + kc_{s,i}^2\delta_i + 3\mathcal{H}c^2_{s,i}u_i \, ,
\end{aligned} \label{Eq:perturb-eqs}
\end{equation}
where $u_i$ is the heat flux and $h$ is the trace of the metric perturbations in the synchronous gauge. The effective sound speed and the conformal Hubble function are defined as
\begin{align}
    c^2_{s, i} \equiv \dfrac{k^2/(4m_i^2a^2)}{1 + k^2/(4m_i^2a^2)}, \quad \mathcal{H} \equiv \dfrac{a'}{a} = aH \, .
\end{align}
We note that Eqs.~\eqref{Eq:perturb-eqs} only show the effective description of axion perturbations when the axion oscillations become much faster than the Hubble timescale. In practice, the exact and effective treatment are combined in our calculations for optimal speed and accuracy, following the procedure in \cite{Hlozek:2014lca, Luu:2021yhl}.

At the first glance, the linear equations of each field in \eqref{Eq:perturb-eqs} seem independent of each other. They are, however, gravitationally coupled to each other via the ``metric" $h$ that are governed by Einstein equations. Consequently, perturbations of the sub-dominant field are modulated by the dominant one on all scales. Fig.~\ref{Fig:ini_spec} (left panels) clearly illustrates this feature in the power spectra of the 2FDM fields. In the 2FDM model (defined in the main text), as $\psi_2$ perturbations accumulate earlier to form potential wells that attract $\psi_1$, the spectrum of $\psi_1$ closely traces that of $\psi_2$ on all scales. On the other hand, these spectra are obviously separate in the single-field FDM model, i.e, the curve of the lighter field ($10^{-22}~{\rm eV}$) has a higher cut-off scale (lower $k$) than to the one of the heavier field ($5 \times 10^{-22}~{\rm eV}$).

Fig.~\ref{Fig:ini_spec} (right panels) also shows that the initial ``clumpiness" of the 2FDM total field is distinguishable from that of the FDM field. If $\psi_1$ and $\psi_2$ were initially set up with the FDM (instead of 2FDM) power spectra in the main text, we would have observed a significantly lower concentration of $\psi_1$ in every halo, which might weaken the argument for halo diversity. It is, thus, crucial that initial conditions in the 2FDM model must be properly inferred from 2FDM linear perturbations.

\renewcommand{\thefigure}{B}
\begin{figure*}
    \centering
    \includegraphics[scale=0.5]{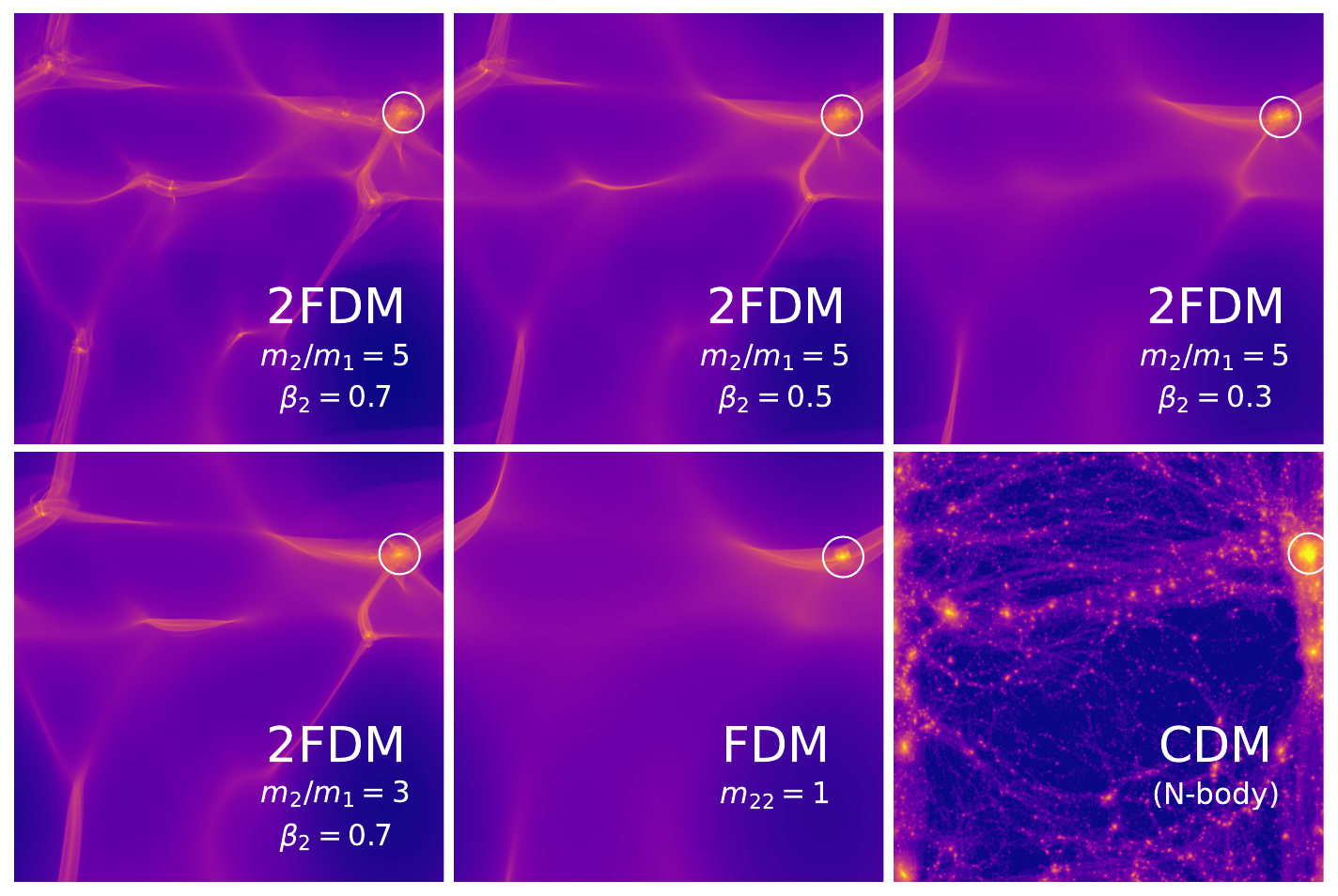} \includegraphics[scale=0.57]{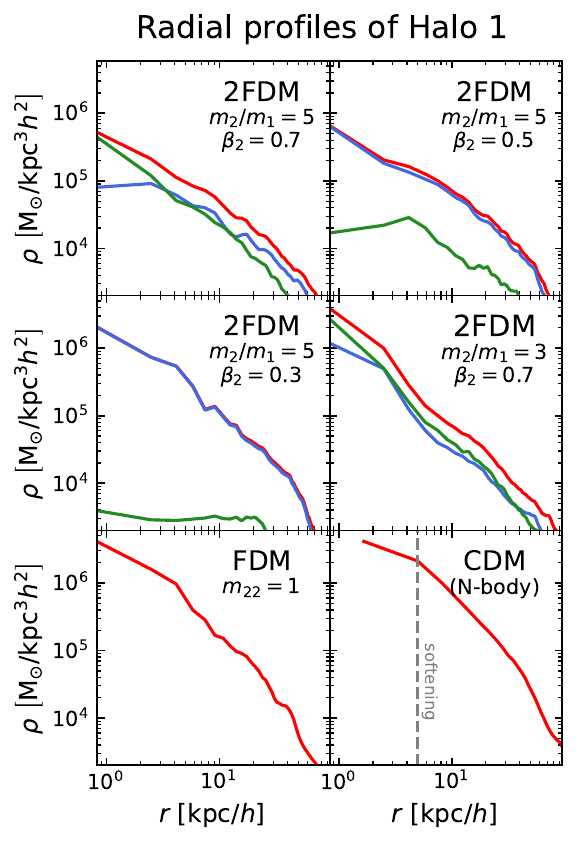} \\
    \caption{(Left) Projected densities of the (total) field in several dark matter models at $z = 5.9$. Input parameters are indicated in each panel. Initial conditions are calculated with the same phases (same random seed) but with different power spectra. The colored density scale is the same as in Fig.~\ref{Fig:proj_evolve}. The 2FDM and FDM simulations shown here have a spectral resolution of $1024^3$, except for the CDM simulation with an N-body resolution of $512^3$ particles. The CDM particles are evolved with \texttt{AREPO}~\cite{Springel:2010, Pakmor:2015ana, Weinberger:2019tbd}. Halo 1 is marked by the white circle in each panel. (Right) The radial profiles of Halo 1 in each dark matter model at $z = 5.9$. The total field, $\psi_1$ and $\psi_2$ correspond to the red, blue and green curves. The FDM and CDM model have only one dark matter component, hence only one density profile (red) plotted. The vertical line (dashed grey) in the last panel denotes a radius equal to three times of the softening length in the CDM simulation.}
    \label{Fig:dens_proj_other}
\end{figure*}

\section{B. Other scenarios}

In the main text, our discussions revolve around a specific framework of the 2FDM model with $m_2/m_1 = 5$ and $\beta_2 = 0.7$. If this scenario is treated as the ``fiducial" model, a natural question is whether halo diversity can be realized in a more generic model. As such, we have performed other simulations (with a resolution of $1024^3$) to examine the impacts of input parameters on the 2FDM cosmology. Fig.~\ref{Fig:dens_proj_other} (left panels) compares structure formation in these scenarios at $z = 5.9$ (due to a lower resolution). By convention we always keep the mass of $\psi_1$ constant at $m_1 = 10^{-22}~{\rm eV}$ and only change $m_2$ when varying $m_2/m_1$.

When the abundance of $\psi_2$ decreases, as for the case of $\beta_2 = 0.5$ or $\beta_2 = 0.3$ with a fixed $m_2/m_1 = 5$, the halo number reduces considerably. We observe three haloes formed in the former and only one formed in the latter case at $z = 5.9$ (compared to eight haloes in the fiducal model) because the initial matter spectrum is more suppressed with more $\psi_1$ in the dark matter budget. Among the haloes that already form, the central region is completely dominated by a soliton of $\psi_1$, {\it e.g.}, see the radial profiles of Halo 1 in Fig.~\ref{Fig:dens_proj_other} (right panels). Compared to the single-field FDM model, these 2FDM models still have an enhanced small-scale density distribution, but the intrinsic halo structures are almost identical, {\it i.e.}, only $\psi_1$-dominated solitons would be observed.

On the other hand, in the model with a lower mass ratio $m_2/m_1 = 3$ and a fixed $\beta_2 = 0.7$, we only find the formation of haloes with nested solitons, which can be seen in Halo 1 but also in two other haloes not shown here. This result seems consistent with what was found by \cite{Huang:2022ffc} in an equivalent set up using the adaptive mesh refinement method. Although the nested soliton is the most unique signature of the 2FDM model, its presence in every halo does not explain the diversity of dwarf galaxies. This is the reason why we need a sufficiently large mass difference between the two species for halo diversity.

There are certainly other aspects that we are unable to study comprehensively due to the limited resolution. Firstly, it is straightforward to notice that the model with $m_2/m_1 = 5$ and $\beta_2 = 0.5$ looks (almost) statistically identical to the one with $m_2/m_1 = 3$ and $\beta_2 = 0.7$ on large scales, which may imply a mass-abundance degeneracy. If that is the case, a 2FDM model defined by $\beta_2$ and $\alpha_2 \equiv m_2/m_1$ is analogous to another model with $\beta'_2 < \beta_2$ and $\alpha'_2 > \alpha_2$, and vice versa. Secondly, it seems halo diversity can only be archived by a combination of a large mass hierarchy and an appropriate density ratio. In case $\beta_2$ is too low or too high, one population of FDM would dominate all dark matter haloes at late times. In our simulations, the threshold of $\beta_2$ seems to fall within $0.6$--$0.7$ for $m_2/m_1 = 5$. From a naive extrapolation based on the mass-abundance degeneracy above, a higher-mass $\psi_2$ would require less $\beta_2$ for halo diversity. In other words, $\psi_1$ would not dominate and suppress the soliton formation of $\psi_2$ at $\beta_2 < 0.7$ thanks to an earlier accumulation of $\psi_2$ in every halo, which also agrees with what we found from idealized simulations in the previous study~\cite{Luu:2023dmi}.

\bibliography{reference}

\end{document}